\documentclass[aps, prd, twocolumn, superscriptaddress, nofootinbib, preprintnumbers]{revtex4}
\usepackage[dvipdfmx,hiresbb]{graphicx} 
\usepackage{graphicx}
\usepackage{amsmath,amssymb}
\usepackage{ulem} 
\usepackage{epsf}
\usepackage{color}   
\usepackage{here}
\allowdisplaybreaks[1]

\newcommand{\nn}{\nonumber\\}

\newcommand{\be}{\begin{equation}}
\newcommand{\e}{\end{equation}}
\newcommand{\aln}[1]{\begin{align}#1\end{align}}

\begin{document}
\preprint{KEK-TH-2019, 
KUNS-2711
}
\title{
Hillclimbing saddle point inflation  
}  
\date{\today}

\author{Kiyoharu Kawana}
\affiliation{
KEK Theory Center, IPNS, Ibaraki 305-0801, Japan
}
\author{Katsuta Sakai}
\affiliation{
Department of Physics, Kyoto University, Kyoto 606-8502, Japan
}

\begin{abstract}
Recently a new inflationary scenario was proposed in \cite{Jinno:2017jxc} which can be applicable to an inflaton having multiple vacua. 
In this letter, we consider a more general situation where 
the inflaton potential has a (UV) saddle point around the Planck scale.  
This class of models can be regarded as a natural generalization of the hillclimbing Higgs inflation \cite{Jinno:2017lun}. 
\end{abstract}

\maketitle

The Standard Model (SM) of particle physics is the most successful theory  that describes physics below the TeV scale. 
The observed Higgs mass $\sim 125$GeV indicates that the SM can be safely interpolated up to the Planck scale without any divergence or instability. 
Furthermore, the observed Higgs quartic coupling $\lambda\sim 0.12$ also shows an interesting behavior of the Higgs potential around the Planck scale $M_{pl}^{}$; The potential can have another degenerate minimum around that scale. 
The origin of this behavior comes from the fact that $\lambda$ and its beta function $\beta_\lambda^{}$ can simultaneously vanish around $M_{pl}^{}$. 
This is called the Multiple point criticality principle and it is surprising that the Higgs mass was predicted to be around $130$GeV about 20 years ago based on this principle \cite{Froggatt:1995rt}. \\
 \indent Various phenomenological and theoretical studies of such a degenerate vacuum have been done so far \cite{Holthausen:2011aa,Bezrukov:2012sa,Degrassi:2012ry,Nielsen:2012pu,Buttazzo:2013uya}. 
One of them is the Higgs inflation with a non minimal coupling $\xi \phi^2 R/M_{pl}^2$ \cite{Bezrukov:2007ep}. 
When this scenario was proposed, it was argued that we need large $\xi\sim 10^5$ in order to obtain the successful inflationary predictions of the cosmic microwave background (CMB). 
However, the criticality of the Higgs potential makes it possible to realize the inflation even if $\xi$ is relatively small $\sim {\cal{O}}(10)$ by using small but nonzero $\lambda\sim 10^{-6}$ around $M_{pl}^{}$. 
See \cite{Hamada:2014wna} for the detailed analyses. \\
\indent Although the SM criticality can help the realization of the Higgs inflation, it is difficult to realize the MPP simultaneously because the latter requires $\lambda=0$ around the Planck scale 
and we can no longer maintain the monotonicity of the Higgs potential above the scale $\sim M_{pl}^{}/\sqrt{\xi}$.   
Recently, a new inflationary scenario was proposed in \cite{Jinno:2017jxc} which enables an inflation even if the inflaton potential has multiple degenerate vacua. 
Then, the authors applied it to the SM Higgs and showed that it is actually possible to obtain a successful inflation while satisfying the MPP \cite{Jinno:2017lun}. 
In those papers, the authors studied a few cases such that the inflaton potential behaves as a quadratic potential around another potential minimum. 
Although the inflationary predictions of this scenario does not strongly depend on the details of the inflaton potential such as the coefficients of the Taylor expansion, they can depend on the leading exponent of the (Jordan-frame) potential and the choice of the conformal factor. 
In this letter, we generalize their works to the cases where the inflaton potential has a saddle point around the Planck scale.   
Our study is meaningful from the point of view of the MPP because this situation can be understood as a natural generalization of this principle.    
Although some fine-tunings are needed in order to realize a saddle point, some theoretical studies \cite{Hamada:2014ofa,Hamada:2014xra,Hamada:2015dja,Kawana:2016tkw} suggest that we can naturally archive such fine-tunings by considering physics beyond ordinary field theory.  

\section{Brief Review of Hillclimbing inflation}
Let us briefly review the hillclimbing inflation. We consider the following action of an inflaton $\phi_J^{}$ in the Jordan-frame: 
\aln{ S=\int d^4x\sqrt{-g_J^{}}\bigg(\frac{M_{pl}}{2}^2\Omega R_J^{}-\frac{K_J^{}}{2}(\partial \phi_J^{})^2-V_J^{}(\phi_J^{})\bigg), 
}
where $(\partial \phi_J^{})^2=g_J^{\mu\nu}\partial_\mu^{}\phi_J^{}\partial_\nu^{}\phi_J^{}$. 
If we identify $\phi_J^{}$ as the Higgs, the usual Higgs potential corresponds to $V_J^{}(\phi_J^{})$ in this framework. 
%
%
Then, by doing the Weyl transformation 
\be g_{\mu\nu}^{}=\Omega g_{J\mu\nu}^{},
\e
we have 
\aln{ 
S=\int d^4x\sqrt{-g}\bigg[\frac{M_{pl}^2}{2} R
-\frac{1}{2}\bigg(\frac{K_J^{}}{\Omega}&+\frac{3}{2}\left(M_{pl}^{}\frac{\partial\ln\Omega}{\partial\phi_J^{}}\right)^2\bigg)(\partial\phi_J^{})^2
\nn
&-\frac{V_J^{}(\phi_J^{})}{\Omega^2}\bigg],
}
where $R$ is the Ricci scalar in the Einstein-frame and we have neglected the total derivative term. 
Let us now assume that the second term of the kinetic terms dominates.  
In this case, we can regard $\chi\equiv M_{pl}^{}\sqrt{3/2}\ln \Omega$ or $-M_{pl}^{}\sqrt{3/2}\ln \Omega$
 as a fundamental field instead of $\phi_J^{}$. \footnote{
The choice of the sign depends on the region we consider; When we consider $\Omega\geq 1\ (\leq 1)$, we take $\chi=(-)M_{pl}^{}\sqrt{3/2}\ln \Omega$.
}
For example, in the case of the ordinary Higgs inflation, we have
\be 
\Omega(\phi_J^{})=1+\xi\frac{\phi_J^2}{M_{pl}^2},\ V_J^{}(\phi_J^{})=\frac{\lambda \phi_J^4}{4}, 
\e
which leads to the following potential in the Einstein-frame: 
\aln{ V_E^{}(\chi)=\frac{\lambda \phi_J^4}{4\Omega^2}
&=\frac{\lambda M_{pl}^4}{4\xi^2}(1-\Omega^{-1})^2\nn
&\simeq \frac{\lambda M_{pl}^4}{4\xi^2}\left(1-\exp\left(-\sqrt{\frac{2}{3}}\frac{\chi}{M_{pl}^{}}\right)\right)^2,
}
from which we can see that $V_E^{}(\chi)$ becomes exponentially flat when $\chi\gg M_{pl}^{}\Leftrightarrow \Omega\gg 1$. 
See also Ref.\cite{Hamada:2014wna} for more detailed analyses.  

On the other hand, a new possibility has been proposed in Ref.\cite{Jinno:2017jxc}, where it is shown that we can also consider the $\Omega\ll1$ region 
instead of $\Omega\gg 1$. 
In this case, because $V_E^{}$ is given by $ V_E^{}=V_J^{}/\Omega^2$, $V_J^{}$ needs to behave as
\be V_J^{}=V_0^{}\Omega^2\left(1+\cdots\right)
\e
around $\Omega=0$ in order to realize the inflationary era, i.e. $H=\dot{a}/a=const$. 
Because the conformal factor $\Omega$ should approaches one after inflation, the inflaton {\it climbs up} the Jordan-frame potential. 
%
This is the reason why the authors of Ref.\cite{Jinno:2017jxc} call this scenario "{\it Hillclimbing (Higgs) inflation}". 
Let us briefly summarize the inflationary predictions of this scenario. By expanding the Jordan-frame potential $V_J^{}$ as a function of $\Omega$ 
\be V_J^{}=V_0^{}\Omega^2(1+\sum_{m\geq n} \eta_m^{}\Omega^m),
\e
we obtain 
\aln{ &\epsilon=\frac{M_{pl}^2}{2}\left(\frac{V'}{V}\right)^2\simeq 
\frac{1}{3}\left(\sum_m\eta_m^{}m\Omega^{m}\right)^2,
\\
&\eta=M_{pl}^2\frac{V''}{V}\simeq -\frac{2}{3}\sum_m\eta_m^{}m^2\Omega^m,
}
where the prime represents the derivative with respect to $\chi$ and we have used the relation $\chi=\sqrt{3/2}\ln\Omega$. 
Furthermore, we can relate the initial value of $\Omega$ to the $e$-folding number $N$:
\aln{ N=\int dt H=\frac{1}{M_{pl}^2}\int d\chi\frac{V}{\frac{\partial V}{\partial\chi}}\simeq \frac{3}{2\eta_n^{}n^2}\frac{1}{\Omega_{ini}^n}.
}
From those equations, we obtain the following inflationary predictions:
\be n_s^{}=1-6\epsilon+2\eta\simeq 1-\frac{2}{N},\ r=16\epsilon=\frac{12}{n^2 N^2}. 
\label{eq: observation of hill}
\e
Note that both of them do not depend on the details of the inflaton potential such as its coefficients $\eta_n^{}$'s. 
This is the similar behavior of the $\xi$ or $\alpha$ attractor \cite{Galante:2014ifa,Kallosh:2013yoa,Kallosh:2014rga}. 
%
However, the leading exponent $n$ depends on a specific model we consider and the choice of the conformal factor. 
In the following, we consider the hillclimbing inflation around a (UV) saddle point of an inflaton potential.  
\section{Hillclimbing Saddle point inflation}
\begin{figure}
\begin{center}
\includegraphics[width=.50\textwidth]{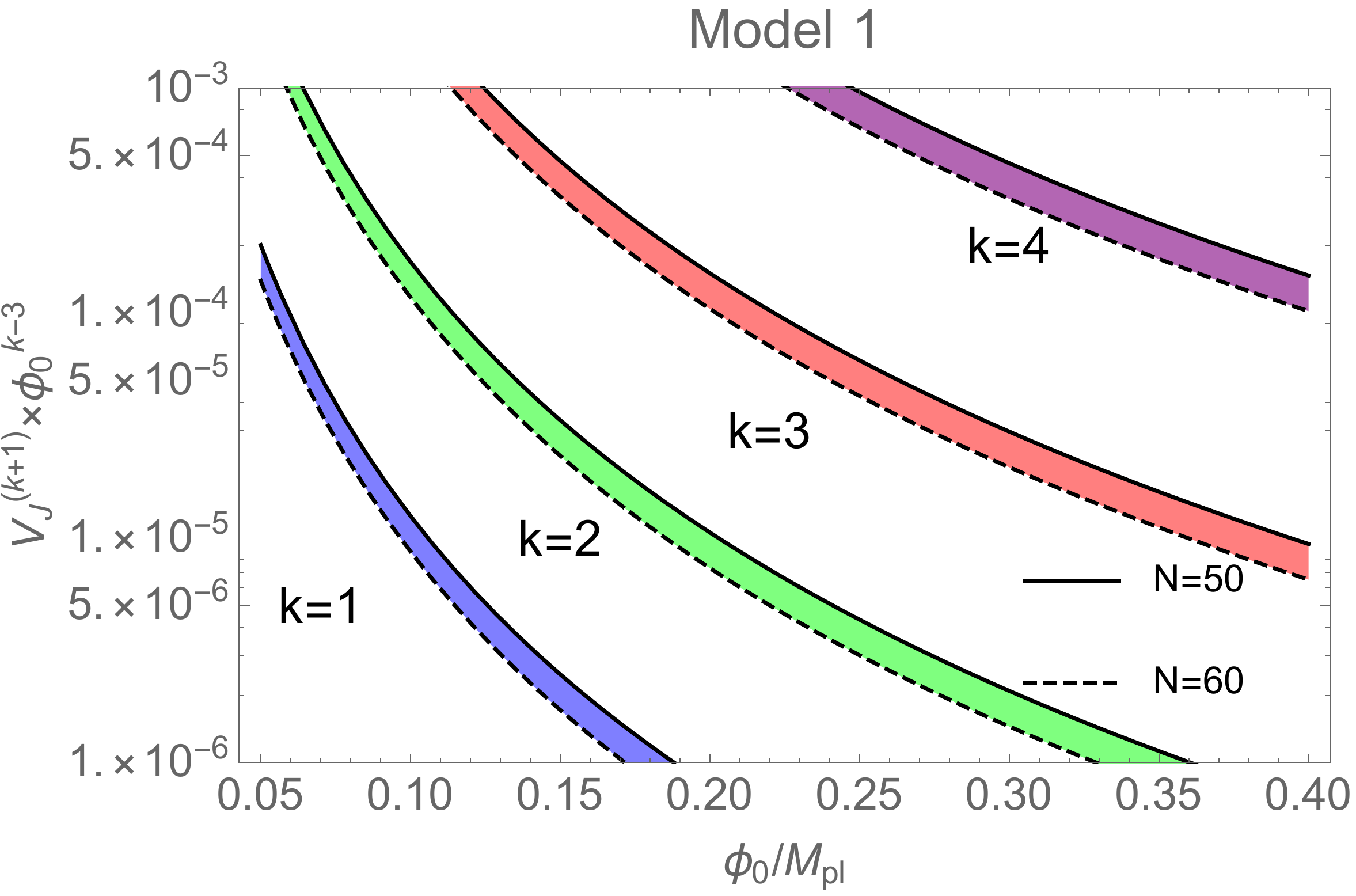}
\includegraphics[width=.50\textwidth]{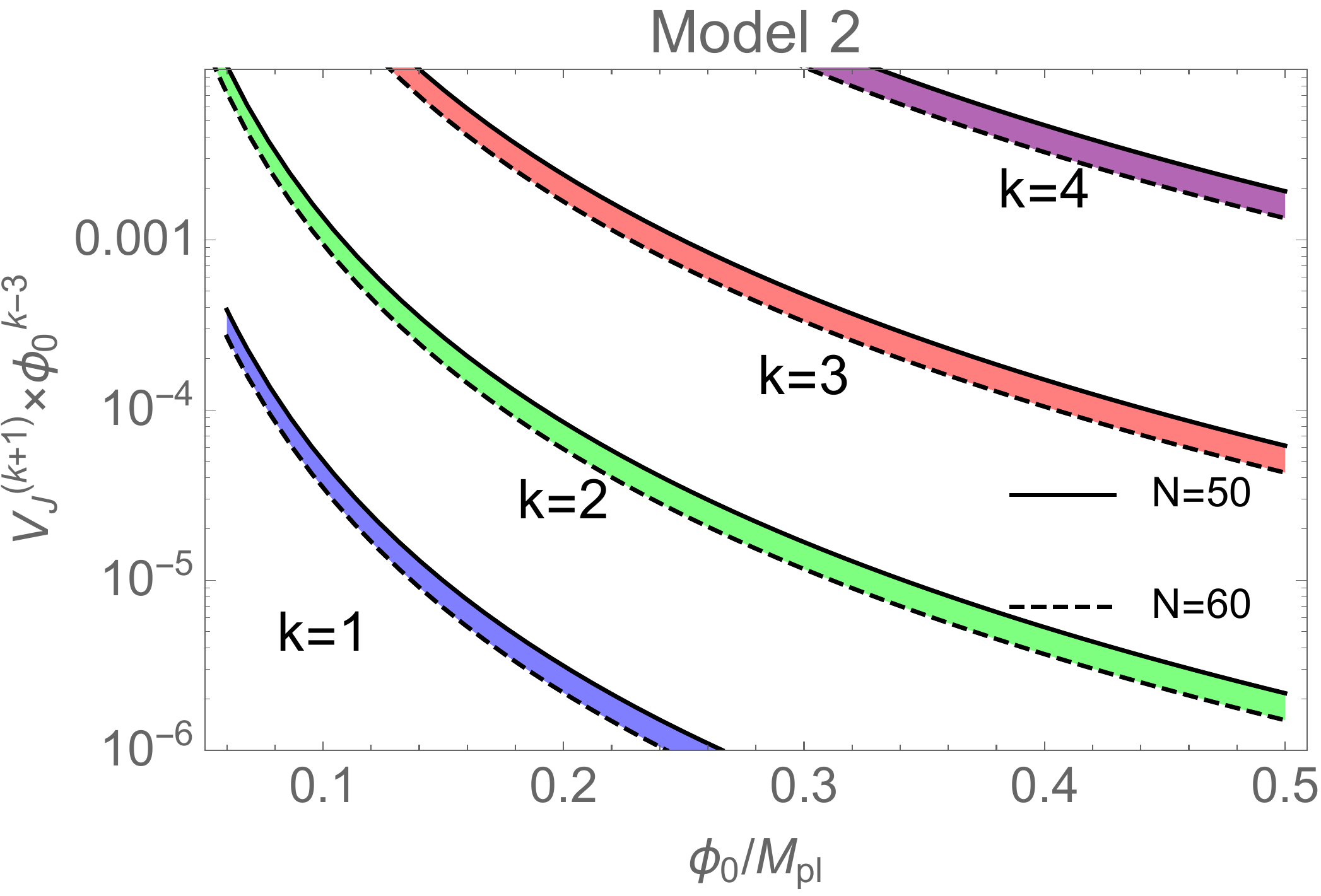}
\caption{The parameter regions that produce the observed value of the scalar perturbation $A_s^{}=2.2\times 10^{-9}$. 
The upper (lower) panel corresponds to Model 1 (2). 
Here, the different color bands represent different $k$'s respectively, and the solid (dashed) lines corresponds to $N=50\ (60)$.  
}
\label{fig:relation}
\end{center}
\end{figure}

\begin{figure}[H]
\begin{center}
\includegraphics[width=.40\textwidth]{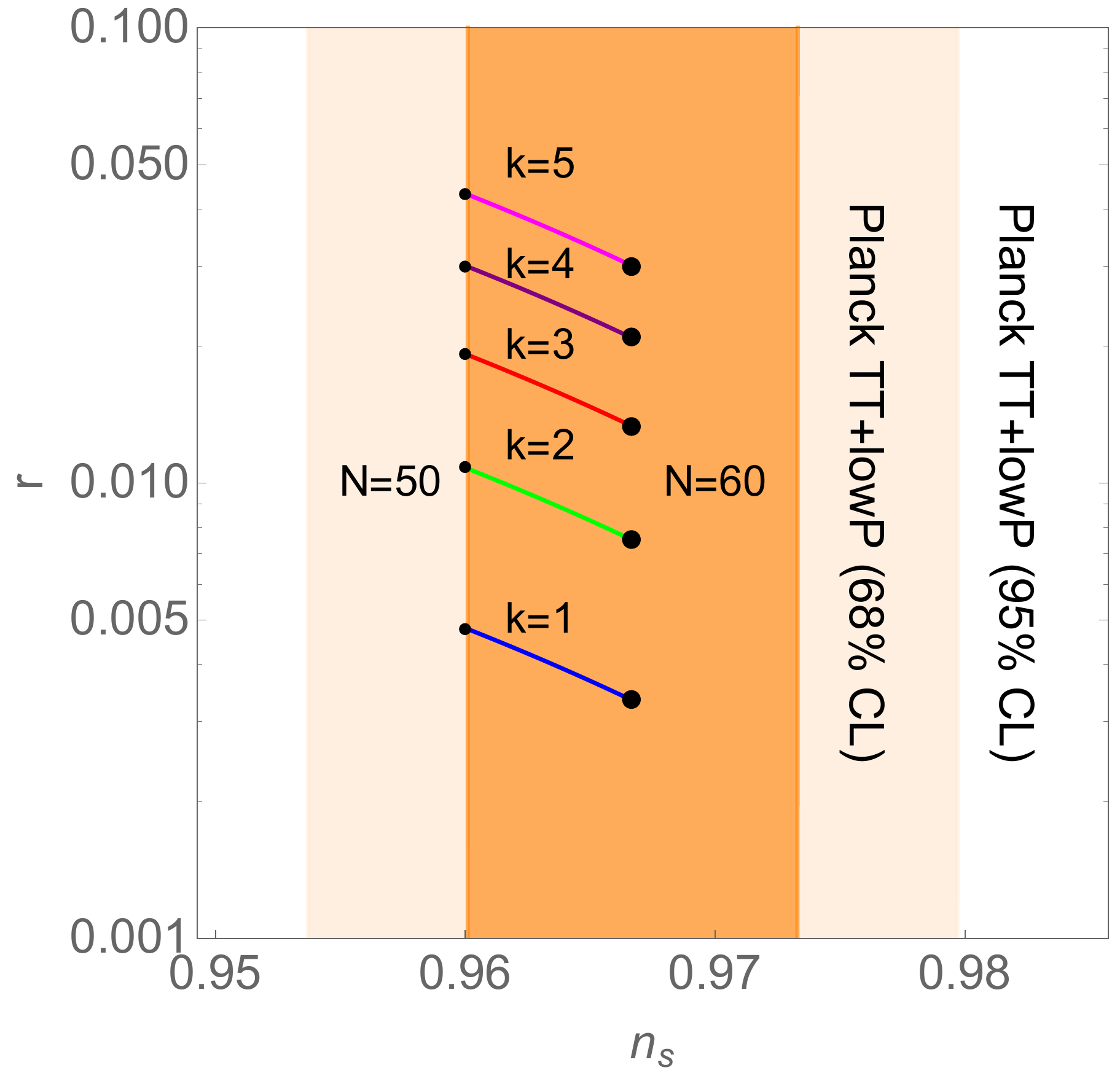}
\caption{The inflationary predictions of the hillclimbing saddle point inflation. 
Here, the different color lines represent different $k$'s and the small (large) dots correspond to $N=50\ (60)$.
}
\label{fig:prediction}
\end{center}
\end{figure}

Let us now consider a general situation where the Jordan-frame potential has a saddle point $\phi_0^{}$ around the Planck scale:
\be V_J^{}(\phi_0^{})=0,\ V^{(1)}_J(\phi_0^{})=0,\ V^{(2)}_J(\phi_0^{})=0,\ \cdots,\ V^{(k)}_J(\phi_0^{})=0
\label{eq: saddle point condition}
\e 
with $V^{(i)}_J$ denoting the $i$-th derivative of $V_J^{}$. 
In the following, we assume 
\be \begin{cases}V_J^{(k+1)}(\phi_0^{}) >0 &\text{for odd }k,
\\
V_J^{(k+1)}(\phi_0^{})<0 &\text{for even }k,
\\
V_J^{(k+2)}(\phi_0^{})\neq 0
\end{cases}
\e
in order to realize a positive vacuum energy in $\phi_J^{}\leq \phi_0^{}$.\footnote{The third assumption is not necessary for our present set up.  
We can also consider a more general situation such that $V_J^{(k+1)}(\phi_0^{})\neq 0,\ V_J^{(k+2)}(\phi_0^{})=0,\ \cdots, V_J^{(k+m)}(\phi_0^{})=0, V_J^{(k+m+1)}(\phi_0^{})\neq 0$. 
}  
\begin{figure}[t]
\begin{center}
\includegraphics[width=.40\textwidth]{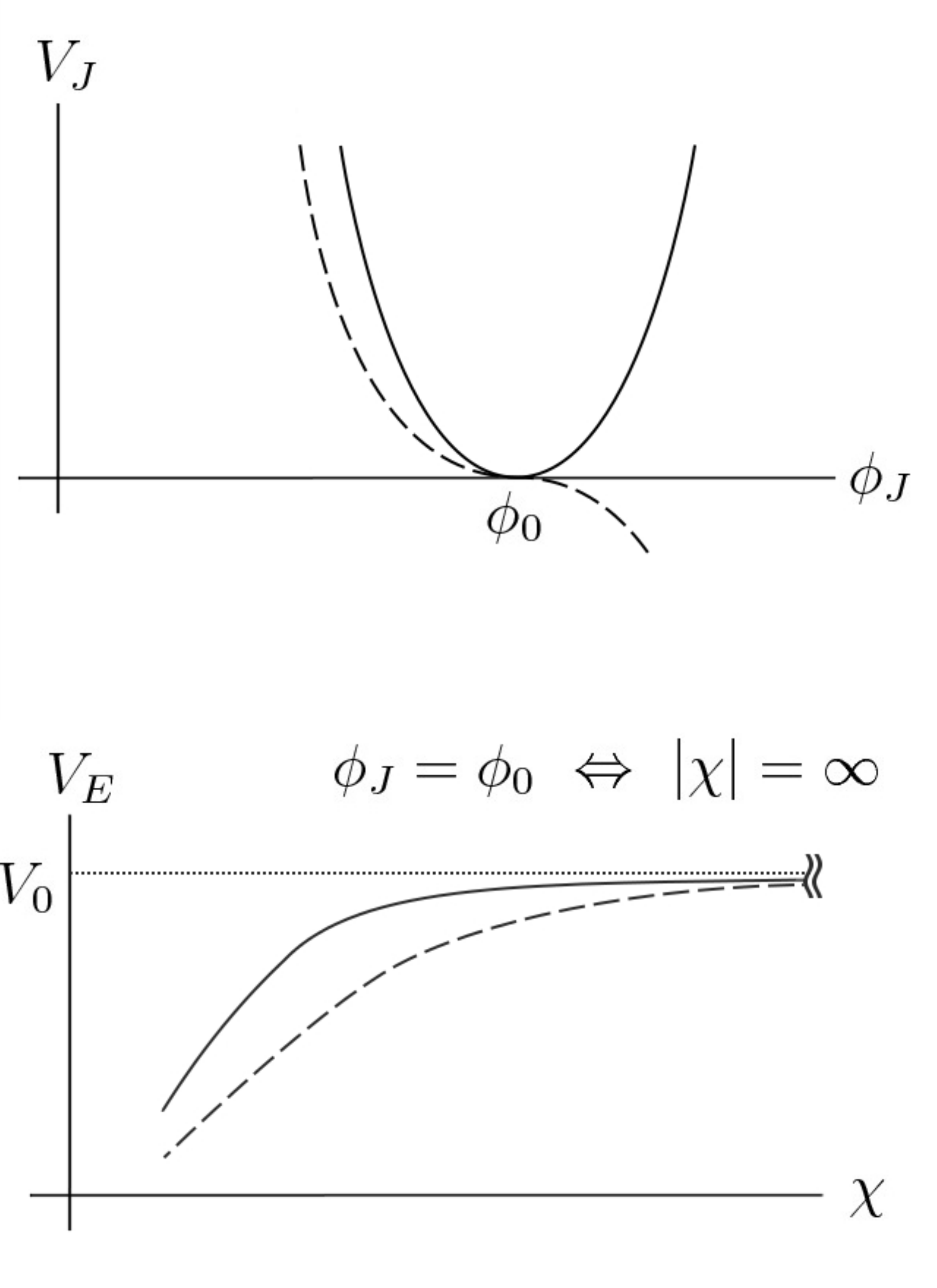}
\caption{Upper (Lower): A schematic behavior of the Jordan (Einstein)-frame potential around the saddle point $\phi_0^{}$ ($\chi=\infty$). 
Here, the solid (dashed) contour corresponds to $k=$odd (even).
}
\label{fig:potential}
\end{center}
\end{figure}
This is schematically shown in the upper panel of Fig.\ref{fig:potential}.
In this case, we can expand $V_J^{}$ around $\phi_0^{}$ as
\aln{ V_J^{}(\phi_J^{})&=\frac{V_J^{(k+1)}}{(k+1)!}(\phi_J^{}-\phi_0^{})^{k+1}+\frac{V_J^{(k+2)}}{(k+2)!}(\phi_J^{}-\phi_0^{})^{k+2}
\nn 
&=\frac{|V_J^{(k+1)}|\phi_0^{k+1}}{(k+1)!}
\left(1-\frac{\phi_J^{}}{\phi_0^{}}\right)^{k+1}
\nn
&\times \bigg(1+v_1^{(k+2)}\left(\frac{\phi_J^{}}{\phi_0^{}}-1\right)+v_2^{(k+3)}\left(\frac{\phi_J^{}}{\phi_0^{}}-1\right)^2\bigg),
\label{eq: expansion of inflaton potential}
}
where
\aln{v_1^{(k+2)}=\frac{\phi_0^{}V_{J}^{(k+2)}}{(k+2)V_{J}^{(k+1)}},\ v_2^{(k+3)}=\frac{\phi_0^{}V_{J}^{(k+3)}}{(k+2)(k+3)V_{J}^{(k+1)}}.
}
As for the conformal factor $\Omega$, we can consider various possibilities:\footnote{
In this letter, we assume that the conformal factor $\Omega$ also becomes zero at a saddle point of $V_J^{}$. 
This fine-tuning might also be explained by some new physics \cite{Hamada:2014ofa,Hamada:2014xra,Hamada:2015dja,Kawana:2016tkw}.
}
\aln{ \Omega(\phi_J^{})^2=&\left(1-\frac{\phi_J^{}}{\phi_0^{}}\right)^{k+1}\left(1+\sum_{i\geq 0}\omega_i^{}\left(1-\frac{\phi_J^{}}{\phi_0^{}}\right)^i\right),
\\
& \sum_{i\geq 0}\omega_i^{}=0, 
}
where the second equation guarantees $\Omega(0)=1$. 
In this letter, in order to give some concrete inflationary predictions, we consider the following two models:
\be \Omega=\begin{cases}\left(1-\frac{\phi_J^{2}}{\phi_0^{2}}\right)^{\frac{k+1}{2}} &\text{(Model 1)},
\\
\left(1-\frac{\phi_J^{4}}{\phi_0^{4}}\right)^{\frac{k+1}{2}} &\text{(Model 2)},
\end{cases}
\e
which correspond to Model 1 and Model 2 presented in Ref.\cite{Jinno:2017lun}, respectively. 
In the case of Model 1, the Einstein-frame potential becomes
\aln{&V_E^{}\simeq \frac{|V_J^{(k+1)}|\phi_0^{k+1}}{(k+1)!2^{k+1}}\bigg(1+\left(\frac{k+1}{2}-v_1^{(k+2)}\right)\left(1-\frac{\phi_J^{}}{\phi_0^{}}\right)
\nn
&+\left(v_2^{(k+3)}-\frac{k+1}{2}v_1^{(k+2)}+\frac{(k+1)(k+2)}{8}\right)\left(1-\frac{\phi_J^{}}{\phi_0^{}}\right)^2\bigg)
\nn
&\simeq V_0^{}\left(1+\eta_{\frac{2}{k+1}}^{}\Omega^{\frac{2}{k+1}}+\eta_{\frac{4}{k+1}}^{}\Omega^{\frac{4}{k+1}}\right),
}
where 
\aln{&V_0^{}=\frac{|V_J^{(k+1)}|\phi_0^{k+1}}{(k+1)!2^{k+1}},\ \eta_{\frac{2}{k+1}}^{}=\frac{1}{2}\left(\frac{k+1}{2}-v_1^{(k+2)}\right),
\nn 
& \eta_{\frac{4}{k+1}}^{}=\frac{1}{2^2}\left(v_2^{(k+3)}-\frac{k+1}{2}v_1^{(k+2)}+\frac{(k+1)(k+2)}{8}\right),
\nn
&\hspace{6cm}(\text{Model 1})
}
from which we can see that the resultant leading exponent depends on the coefficients of the Jordan-frame potential. 
\footnote{For example, in the case of the Higgs potential, we have $k=1, v_1^{(k+2)}=3$, which lead to $\eta_{1}^{}=-1$. 
This agrees with the previous study Ref.\cite{Jinno:2017lun}. 
}  
In the lower panel of Fig.\ref{fig:potential}, we schematically show the  Einstein-frame potential $V_E^{}$. 
Here note that the saddle point $\phi_0^{}$ corresponds to $\chi=\infty$ because of the relation $\chi=-M_{pl}^{}\sqrt{3/2}\ln\Omega$.   
Here, the solid (dashed) contour corresponds to $k=$odd (even).
%
In the case of Model 2, we have 
\aln{&V_0^{}=\frac{|V_J^{(k+1)}|\phi_0^{k+1}}{(k+1)!2^{2(k+1)}},\ \eta_{\frac{2}{k+1}}^{}=\frac{1}{4}\left(\frac{3(k+1)}{2}-v_1^{(k+2)}\right),
\nn 
& \eta_{\frac{4}{k+1}}^{}=\frac{1}{4^2}\left(v_2^{(k+3)}-\frac{3(k+1)}{2}v_1^{(k+2)}+\frac{(k+1)(9k+10)}{8}\right),
\nn
&\hspace{6cm}(\text{Model 2})
}
Thus, both of the models typically give the leading exponent $n=\frac{2}{(k+1)}$ as long as we do not require a fine-tuning of the coefficients.\footnote{
If we consider general $V_J^{}$ and $\Omega$, the coefficients $\eta_{2i/(k+1)}$'s are simple polynomials of $(v_i^{(k+i+1)},\omega_i^{})$, and it is possible to eliminate some of the first $\eta_{2i/(k+1)}$'s by choosing specific values of those parameters. Then, the leading exponent can be $n=\frac{2l}{k+1}$ with arbitrary $l$. 
The Model 2 of the hillclimbing Higgs inflation Ref.\cite{Jinno:2017lun} is such a case.
}
As a result, the tensor-to-scalar ratio becomes larger when we increase $k$. 
Note that, in this framework, the coefficient of the leading term in the potential must be negative, $\eta_\frac{2}{k+1}<0$, which enables $\chi$ to roll down it. 
Furthermore, the potential height $V_0^{}$ is also constrained by the curvature perturbation 
\be A_s^{}=\frac{V_0^{}}{24\pi^2\epsilon M_{pl}^4}=2.2\times 10^{-9}\propto \frac{V_J^{(k+1)}(\phi_0^{})\phi_0^{k+1}}{M_{pl}^4}.\label{eq: As}
\e
In Fig.\ref{fig:relation}, we plot the parameter regions obtained from Eq.(\ref{eq: As}). 
Here, the $(k+1)$-th derivative $V_J^{(k+1)}(\phi_0^{})$ is normalized by $\phi_0^{k-3}$, and each bands corresponds to each $k$'s. 
The solid (dashed) contours represent $N=50\ (60)$.       
In Fig.\ref{fig:prediction}, we also show the inflationary predictions obtained from the analytic formulas Eq.(\ref{eq: observation of hill}). 
Here, the different color lines represent different $k$'s and the small (large) dots correspond to $N=50\ (60)$. 
Note that $n_s^{}$ does not change within this analytic formula because it only depends on the $e$-folding $N$. 
As is already mentioned in Ref.\cite{Jinno:2017lun}, the higher order terms of the inflaton potential can have slightly large contributions to the inflationary dynamics, and numerical studies are necessary in order to give more precise predictions. 
This is left for future investigations.

\section{Conclusion}
In this letter, we have applied the idea of the hillclimbing inflation to the models where the inflaton potential has a saddle point around the Planck scale and shown that it is possible to archive a successful inflation. 
A notable feature of this class of models is that the leading exponent of the Jordan-frame potential as a function of the conformal factor is typically given by $2/(k+1)$, which leads to a large tensor-to-scalar ratio.  
Although we have just concentrated on a saddle point of the inflaton potential, we can also consider various realizations of the hillclimbing inflation by using a variety of $V_J^{}$ and $\Omega$.   
So it might be interesting to investigate such possibilities and construct a phenomenological model that can realize a successful inflation.  
\\
\\
%

%
%

\subsection*{Acknowledgement}
We thank H. Kawai and R.Jinno for valuable comments. 
The work of KK (KS) is supported by the Grant-in-Aid for JSPS Research Fellow, Grant Number 17J03848 (17J02185).

\end{document}